\documentclass{article}

\PassOptionsToPackage{numbers, compress}{natbib}

\usepackage[final]{neurips_2024}
\usepackage{amsmath}
\usepackage{mathtools}
\usepackage{booktabs}
\usepackage{multirow}
\usepackage{graphicx}
\usepackage[table,xcdraw]{xcolor}
\usepackage{colortbl} 
\usepackage{caption}
\usepackage{subcaption}
\usepackage{wrapfig}
\usepackage{longtable}

\usepackage[utf8]{inputenc} 
\usepackage[T1]{fontenc}    
\usepackage{hyperref}       
\usepackage{url}            
\usepackage{booktabs}       
\usepackage{amsfonts}       
\usepackage{nicefrac}       
\usepackage{microtype}      
\usepackage{xcolor}         

\AtEndPreamble{
    \usepackage[capitalize]{cleveref}
    \crefname{section}{Sec.}{Secs.}
    \Crefname{section}{Section}{Sections}
    \crefname{table}{Tab.}{Tabs.}
    \Crefname{table}{Table}{Tables}
}

\definecolor{colorf}{HTML}{FF9396}
\definecolor{colors}{HTML}{FFC991}
\definecolor{colort}{HTML}{FFF6A9}
\newcommand{\boxcolorf}[1]{%
  \begingroup\setlength{\fboxsep}{1pt}%
  \colorbox{colorf}{\hspace*{2pt}\vphantom{Ay}#1\hspace*{2pt}}%
  \endgroup
}
\newcommand{\boxcolors}[1]{%
  \begingroup\setlength{\fboxsep}{1pt}%
  \colorbox{colors}{\hspace*{2pt}\vphantom{Ay}#1\hspace*{2pt}}%
  \endgroup
}
\newcommand{\boxcolort}[1]{%
  \begingroup\setlength{\fboxsep}{1pt}%
  \colorbox{colort}{\hspace*{2pt}\vphantom{Ay}#1\hspace*{2pt}}%
  \endgroup
}

\newcommand{\hatbf}[1]{\hat{\mathbf{#1}}}
\newcommand{\tildebf}[1]{\Tilde{\mathbf{#1}}}

\title{R$^2$-Gaussian: Rectifying Radiative Gaussian Splatting for Tomographic Reconstruction}

\author{%
Ruyi Zha$^{1}$ \quad Tao Jun Lin$^{1}$ \quad Yuanhao Cai$^{2,}$\thanks{\footnotesize Yuanhao Cai is the corresponding author.} \quad Jiwen Cao$^1$ \\
 \textbf{Yanhao Zhang}$^3$ \quad \textbf{Hongdong Li}$^1$\\\\
$^1$The Australian National University \quad $^2$Johns Hopkins University \\
$^3$Robotics Institute, University of Technology Sydney\\
\texttt{\{ruyi.zha, taojun.lin, jiwen.cao, hongdong.li\}@anu.edu.au}\\
\texttt{caiyuanhao1998@gmail.com} \quad 
\texttt{yanhao.zhang@uts.edu.au}\\
}

\begin{document}

\maketitle

\begin{abstract}
3D Gaussian splatting (3DGS) has shown promising results in image rendering and surface reconstruction. However, its potential in volumetric reconstruction tasks, such as X-ray computed tomography, remains under-explored. This paper introduces R$^2$-Gaussian, the first 3DGS-based framework for sparse-view tomographic reconstruction. By carefully deriving X-ray rasterization functions, we discover a previously unknown \emph{integration bias} in the standard 3DGS formulation, which hampers accurate volume retrieval. To address this issue, we propose a novel rectification technique via refactoring the projection from 3D to 2D Gaussians. Our new method presents three key innovations: (1) introducing tailored Gaussian kernels, (2) extending rasterization to X-ray imaging, and (3) developing a CUDA-based differentiable voxelizer. Experiments on synthetic and real-world datasets demonstrate that our method outperforms state-of-the-art approaches in accuracy and efficiency. Crucially, it delivers high-quality results in 4 minutes, which is 12$\times$ faster than NeRF-based methods and on par with traditional algorithms. Code and models are available on the project page \url{https://github.com/Ruyi-Zha/r2_gaussian}.

\end{abstract}

\section{Introduction}
\label{sec: introduction}

Computed tomography (CT) is an essential imaging technique for noninvasively examining the internal structure of objects. Most CT systems use X-rays as the imaging source thanks to their ability to penetrate solid substances~\cite{kak2001principles}. 
During a CT scan, an X-ray machine captures multi-angle 2D projections that measure ray attenuation through the material. As the core of CT, tomographic reconstruction aims to recover the 3D density field of the object from its projections. This task is challenging in two aspects. Firstly, the harmful X-ray radiation limits the acquisition of sufficient and noise-free projections, making reconstruction a complex and ill-posed problem. Secondly, time-sensitive applications like medical diagnosis require algorithms to deliver results promptly.

Existing tomography methods suffer from either suboptimal reconstruction quality or slow processing speed. Traditional CT algorithms~\cite{feldkamp1984practical,andersen1984simultaneous,sidky2008image} deliver results in minutes but induce serious artifacts. Supervised learning-based approaches~\cite{lin2023learning,lin2024c,chung2023solving,liu2023dolce} achieve promising outcomes by learning semantic priors but struggle with out-of-distribution objects. Recently, neural radiance fields (NeRF)~\cite{mildenhall2020nerf} have been applied to tomography and perform well in per-case reconstruction~\cite{zha2022naf, zang2021intratomo, ruckert2022neat,cai2023structure,shen2022nerp}. However, they are very time-consuming ($>30$ minutes) because a huge amount of points have to be sampled for volume rendering.

\begin{figure}[t]
    \centering
    \includegraphics[width=\linewidth]{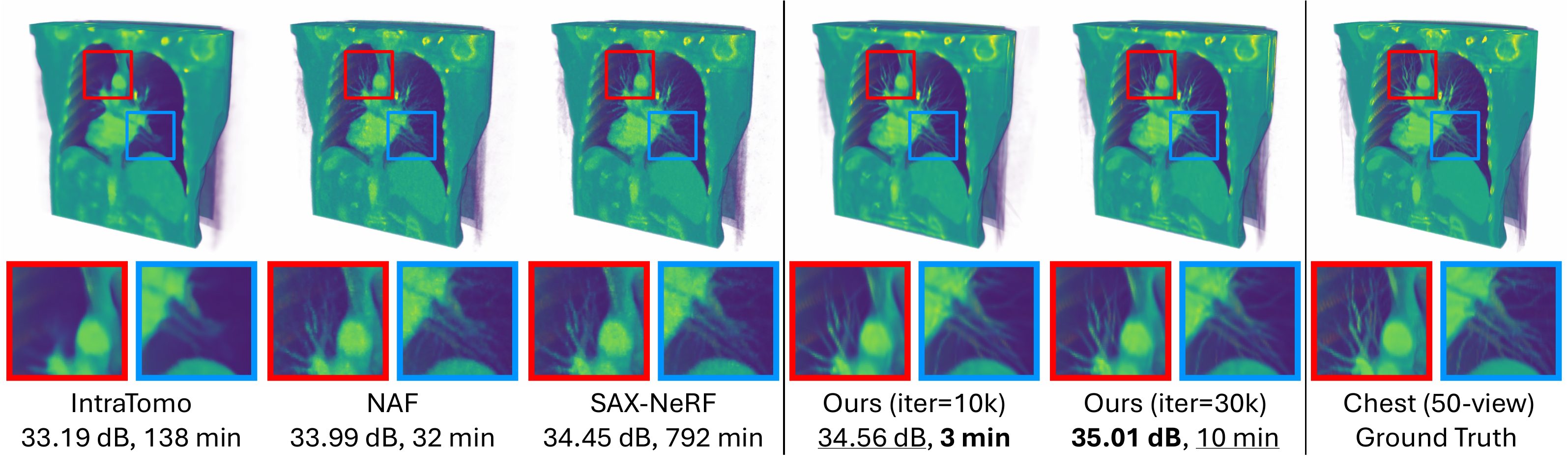}
    \caption{
    We compare our method to state-of-the-art NeRF-based methods (IntraTomo~\cite{zang2021intratomo}, NAF~\cite{zha2022naf}, SAX-NeRF~\cite{cai2023structure}) in terms of visual quality, PSNR (dB), and training time (minute). Our method achieves the highest reconstruction quality and is significantly faster than other methods.}
    \label{fig: cover image}
\end{figure}

Recently, 3D Gaussian splatting (3DGS)~\cite{kerbl20233d} has outperformed NeRF in both quality and efficiency for view synthesis~\cite{Yu2023MipSplatting,scaffoldgs,liang2023gs} and surface reconstruction~\cite{guedon2023sugar, Huang2DGS2024,Yu2024GOF}. However, attempts to apply the 3DGS technique to volumetric reconstruction tasks, such as X-ray tomography, are limited and ineffective. Some concurrent works \cite{cai2024radiative,gao2024ddgs} empirically modify 3DGS for X-ray view synthesis, but they treat it solely as a data augmentation tool for traditional tomography algorithms. To date, there is no 3DGS-based method for direct CT reconstruction.

In this paper, we reveal an inherent \textbf{integration bias} in 3DGS. This bias, despite having a negligible impact on image rendering, critically hampers volumetric reconstruction.  To be more specific, we will show in~\cref{sec: x-ray rasterization} that the standard 3DGS overlooks a covariance-related scaling factor when splatting a 3D Gaussian kernel onto the 2D image plane. This formulation leads to inconsistent volumetric properties queried from different views. Besides the integration bias, there are other challenges in applying 3DGS to tomography, such as the difference between natural light and X-ray imaging and the lack of an effective technique to query volumes from kernels.

We propose R$^2$-Gaussian (Rectified Radiative Gaussians) to extend 3DGS to sparse-view tomographic reconstruction.  R$^2$-Gaussian achieves a bias-free training pipeline with three significant improvements. \textbf{Firstly}, we introduce a novel radiative Gaussian kernel, which acts as a local density field parameterized by central density, position, and covariance.  We initialize Gaussian parameters using the analytical method FDK~\cite{feldkamp1984practical} and optimize them with photometric losses. \textbf{Secondly}, we rectify the 3DGS rasterizer to support X-ray imaging. This is achieved by deriving new X-ray rendering functions and correcting the integration bias for accurate density retrieval. \textbf{Thirdly}, we develop a CUDA-based differentiable voxelizer, which not only extracts 3D volumes from Gaussians but also enables voxel-based regularization during training.  We evaluate R$^2$-Gaussian on both synthetic and real-world datasets.  Extensive experiments demonstrate that our method surpasses state-of-the-art (SOTA) methods within 4 minutes, which is $ 12\times$ faster than the most efficient NeRF-based solution, NAF~\cite{zha2022naf} and comparable to traditional algorithms. It converges to optimal results in 15 minutes, improving PSNR by 0.6 dB compared to SOTA methods. A visual comparison is shown in~\cref{fig: cover image}.

Our contributions can be summarized as follows: (1) We discover a previously unknown integration bias in 3DGS that impedes volumetric reconstruction. (2) We propose the first 3DGS-based tomography framework by introducing new kernels, extending rasterization to X-ray imaging, and developing a differentiable voxelizer. (3) Our method significantly outperforms state-of-the-art methods in both reconstruction quality and training speed, highlighting its practical value.

\section{Related work}
\label{sec: related work}
\paragraph{Tomographic reconstruction} Computed tomography (CT) is widely used for non-intrusive inspection in medicine~\cite{hounsfield1980computed, katsuragawa2007computer}, biology~\cite{donoghue2006synchrotron,luvcic2005structural,kiljunen2015dental}, and industry~\cite{de2014industrial}. Conventional fan-beam CT produces a 3D volume by reconstructing each slice from 1D projection arrays. Recently, the cone-beam scanner has become popular for its fast scanning and high resolution~\cite{scarfe2006clinical}, leading to the demand for 3D tomography, i.e., recovering the volume directly from 2D projection images. Our work focuses on 3D sparse-view reconstruction where less than a hundred projections are captured to reduce radiation exposure. Traditional algorithms are mainly grouped into analytical and iterative methods. Analytical methods like filtered back projection (FBP) and its 3D variant FDK~\cite{feldkamp1984practical} produce results instantly ($<1$ second) by solving the Radon transform and its inverse~\cite{radon1986determination}. However, they introduce serious streak artifacts in sparse-view scenarios. Iterative methods~\cite{andersen1984simultaneous,sidky2008image,manglos1995transmission,sauer1993local} formulate tomography as a maximum-a-posteriori problem and iteratively minimize the energy function with regularizations. They successfully suppress artifacts but take longer time ($<10$ minutes) and lose structure details. Deep learning methods can be categorized as supervised and self-supervised families. Supervised methods learn semantic priors from CT datasets. They then use the trained networks to inpaint projections~\cite{anirudh2018lose,ghani2018deep}, denoise volumes~\cite{chung2023solving,lee2023improving,liu2023dolce,liu2020tomogan} or directly output results~\cite{jin2017deep,ying2019x2ct,adler2018learned,lin2023learning,lin2024c}. Supervised learning methods perform well in cases similar to training sets but suffer from poor generation ability when applied to unseen data. To overcome this limitation, some studies~\cite{zha2022naf, zang2021intratomo, ruckert2022neat,cai2023structure,shen2022nerp} handle tomography in a self-supervised learning fashion. Inspired by NeRF~\cite{mildenhall2020nerf}, they model the density field with coordinate-based networks and optimize them with photometric losses. Although NeRF-based methods excel in per-case reconstruction, they are time-consuming ($>$30 minutes) due to the extensive point sampling in volume rendering. Our work can be put into the self-supervised learning family, but it greatly accelerates the training process and improves reconstruction quality.

\paragraph{3DGS} 3D Gaussian splatting~\cite{kerbl20233d} outperforms NeRF in speed by leveraging highly parallelized rasterization for image rendering. 3DGS represents objects with a set of trainable Gaussian-shaped primitives. It has achieved great success in RGB tasks, including surface reconstruction~\cite{guedon2023sugar, Huang2DGS2024,Yu2024GOF}, dynamic scene modeling~\cite{wu20234d,lin2023gaussian,yang2023deformable}, human avatar~\cite{liu2023humangaussian,li2024animatablegaussians,kocabas2024hugs}, 3D generation~\cite{tang2024dreamgaussian,yi2023gaussiandreamer,chen2023text}, etc. Some concurrent works have extended 3DGS to X-ray imaging. X-Gaussian~\cite{cai2024radiative} modify 3DGS to synthesize novel-view X-ray projections. \citet{gao2024ddgs} improve X-Gaussian by considering complex noise-inducing physical effects. While they produce plausible 2D X-ray projections, they cannot directly extract 3D density volumes from trained Gaussians. Instead, they first augment projections with 3DGS, and then use traditional algorithms such as FDK for CT reconstruction, which is neither efficient nor effective. \citet{li2023sparse} represent the density field with customized Gaussian kernels, but they replace the efficient rasterization with existing CT simulators. In comparison, our work can both rasterize X-ray projections and voxelize density volumes from Gaussians.

\section{Preliminary}
\label{sec: preliminary}

\subsection{X-ray imaging}
\label{sec: x-ray imaging}

\begin{wrapfigure}{r}{0.46\textwidth}
    \centering
    \includegraphics[width=0.46\textwidth]{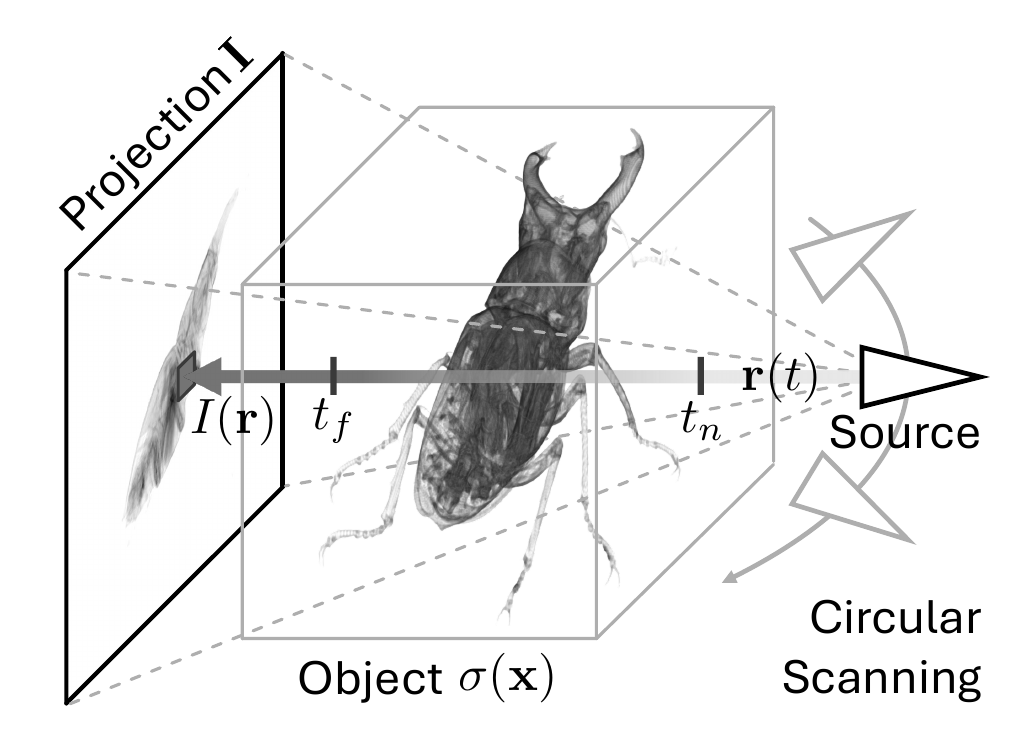}
    \caption{A detection plane captures the attenuation of X-rays emitted from different angles.}
    \label{fig: x-ray imaging}
\end{wrapfigure}

A projection $\mathbf{I}\in \mathbb{R}^{H\times W}$ measures ray attenuation through the material as shown in~\cref{fig: x-ray imaging}. For an X-ray $\mathbf{r}(t) = \mathbf{o} + t\mathbf{d} \in \mathbb{R}^3$ with initial intensity $ I_0$ and path bounds $t_n$ and $t_f$, the corresponding raw pixel value $I'(\mathbf{r})$ is given with the Beer-Lambert Law~\cite{kak2001principles} by: $\smash{I'(\mathbf{r}) =  I_0 \exp(-\int_{t_n}^{t_f} \sigma(\mathbf{r}(t)) \, dt)}$. Here, $\sigma(\mathbf{x})$ is the isotropic density (or attenuation coefficient in physics)  at position $\mathbf{x} \in \mathbb{R}^3$. Tomography typically transforms raw data to the logarithmic space for computational simplicity, i.e.,
\begin{equation}
\label{equ: x-ray imaging}
    I(\mathbf{r}) = \log  I_0 - \log  I'(\mathbf{r}) =\int_{t_n}^{t_f} \sigma(\mathbf{r}(t)) dt,
\end{equation}
where each pixel value $I(\mathbf{r})$ represents the density integral along the ray path. Except otherwise specified, we use the logarithmic projections as inputs. The goal of tomographic reconstruction is to estimate the 3D distribution of $\sigma(\mathbf{x})$, output as a discrete volume, with X-ray projections $\{\mathbf{I}_i\}_{i=1,\cdots,N}$ captured from $N$ different angles. Note that real-world projections contain minor anisotropic physical effects such as Compton scattering. Following previous works~\cite{feldkamp1984practical,andersen1984simultaneous,sidky2008image,zha2022naf}, we do not explicitly model them but treat them as noise during the reconstruction.

\subsection{3D Gaussian splatting}
\label{sec: preliminary 3dgs}
3D Gaussian splatting~\cite{kerbl20233d} models the scene with a set of 3D Gaussian kernels $\mathbb{G}^{3}=\{G^3_{i}\}_{i=1,\cdots,M}$, each parameterized by position, covariance, color, and opacity. A rasterizer $\mathcal{R}$ renders an RGB image $\mathbf{I}_{rgb}\in \mathbb{R}^{H\times W\times3}$ from these Gaussians, formulated as
\begin{equation}
    \mathbf{I}_{rgb} = \mathcal{R}(\mathbb{G}^{3}) = \mathcal{C}\circ \mathcal{P}\circ \mathcal{T}(\mathbb{G}^{3}),
\end{equation}
where $\mathcal{T}$, $\mathcal{P}$, and $\mathcal{C}$ are the transformation, projection, and composition modules, respectively. First, $\mathcal{T}$ transforms the 3D Gaussians into the ray space, aligning viewing rays with the coordinate axis to enhance computational efficiency. The transformed 3D Gaussians are then projected onto the image plane: $\mathbb{G}^2 = \mathcal{P}(\mathbb{G}^3)$. The projected 2D Gaussian retains the same opacity and color as its 3D counterpart but omits the third row and column of position and covariance. An RGB image is then rendered by compositing these 2D Gaussians using alpha-blending~\cite{porter1984compositing}: $\mathbf{I}_{rgb} = \mathcal{C}(\mathbb{G}^2)$. The rasterizer is differentiable, allowing for the optimization of kernel parameters using photometric losses. 3DGS initializes sparse Gaussians with structure-from-motion (SfM) points~\cite{schonberger2016structure}. During training, an adaptive control strategy dynamically densifies Gaussians to improve scene representation.

\section{Method}
\label{sec: method}

In this section, we first introduce radiative Gaussian as a novel object representation in~\cref{sec: radiative gaussians}. Next, we adapt 3DGS to tomography in~\cref{sec: training radiative Gaussians}. Specifically, we derive new rasterization functions and analyze the integration bias of standard 3DGS in~\cref{sec: x-ray rasterization}. We further develop a differentiable voxelizer for volume retrieval in \cref{sec: density voxelization}. The optimization strategy is elaborated in~\cref{sec: optimization}.

\subsection{Representing objects with radiative Gaussians}
\label{sec: radiative gaussians}

\begin{figure}[t]
    \centering
    \includegraphics[width=\linewidth]{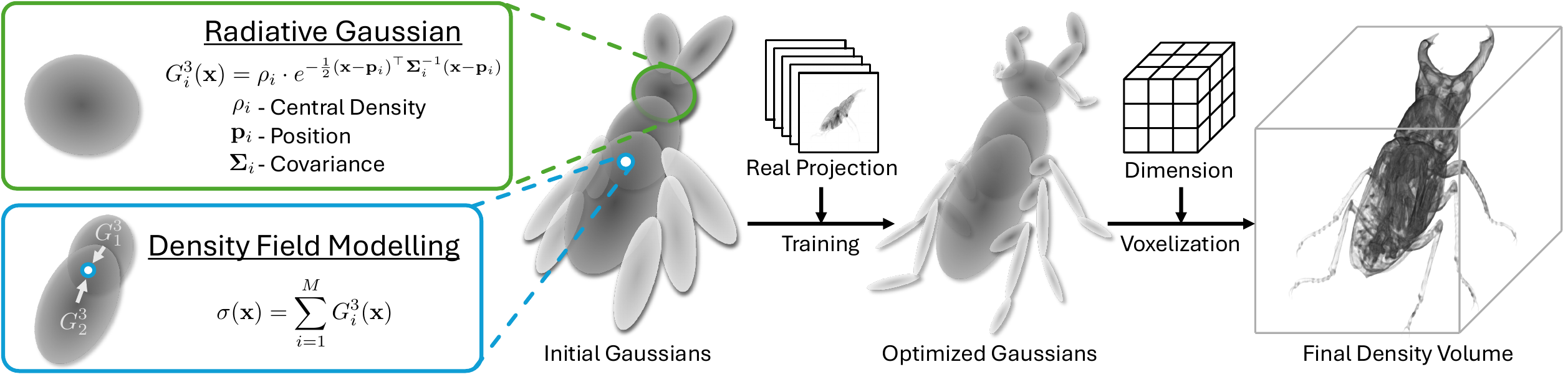}
    \caption{We represent the scanned object as a set of radiative Gaussians. We optimize them using real X-ray projections and finally retrieve the density volume with voxelization.}
    \label{fig: radiative Gaussian}
\end{figure}

As shown in~\cref{fig: radiative Gaussian}, we represent the target object with a group of learnable 3D kernels $\mathbb{G}^{3}=\{G^3_{i}\}_{i=1,\cdots,M}$ that we term as radiative Gaussians. Each kernel ${G^3_{i}}$ defines a local Gaussian-shaped density field, i.e.,

\begin{equation}
\label{equ: one gaussian}
    G^3_{i}(\mathbf{x}|\rho_i, \mathbf{p}_{i}, \mathbf{\Sigma}_{i})=\rho_{i} \cdot \exp \left( -\frac{1}{2}(\mathbf{x}-\mathbf{p}_{i})^\top \mathbf{\Sigma}_{i}^{-1}(\mathbf{x}-\mathbf{p}_{i}) \right),
\end{equation}

where $\rho_{i}$, $\mathbf{p}_{i}\in \mathbb{R}^3$ and $\mathbf{\Sigma}_{i}\in \mathbb{R}^{3\times3}$ are learnable parameters representing central density, position and covariance, respectively. For optimization purposes, we follow~\cite{kerbl20233d} to further decompose the covariance matrix $\mathbf{\Sigma}_{i}$ into the rotation matrix $\mathbf{R}_{i}$ and scale matrix $\mathbf{S}_{i}$: $\mathbf{\Sigma}_{i}=\mathbf{R}_{i}\mathbf{S}_{i}\mathbf{S}_{i}^\top \mathbf{R}_{i}^\top$. The overall density at position $\mathbf{x}\in \mathbb{R}^{3}$ is then computed by summing the density contribution of kernels:
\begin{equation}
\label{equ: density as sum of 3d Gaussians}
    \sigma(\mathbf{x})=\sum_{i=1}^{M} G^3_{i}(\mathbf{x}|\rho_i, \mathbf{p}_{i}, \mathbf{\Sigma}_{i}).
\end{equation}
Compared with standard 3DGS, our kernel formulation removes view-dependent color because X-ray attenuation depends only on isotropic density, as shown in \cref{equ: x-ray imaging}. More importantly, we define the density query function (\cref{equ: density as sum of 3d Gaussians}) for radiative Gaussians, making them useful for both 2D image rendering and 3D volume reconstruction. In contrast, the opacity in 3DGS is empirically designed for RGB rendering, leading to challenges when extracting 3D models such as meshes from Gaussians~\cite{guedon2023sugar, chen2023neusg, Yu2024GOF}. Concurrent work~\cite{li2023sparse} also explores kernel-based representation but uses simplified isotropic Gaussians. Our work employs a general Gaussian distribution, offering more flexibility and precision in modeling complex structures.

\paragraph{Initialization} 3DGS initializes Gaussians with SfM points, which is not applicable to volumetric tomography. Instead, we initialize our radiative Gaussians using preliminary results obtained from the analytical method. Specifically, we use FDK~\cite{feldkamp1984practical} to reconstruct a low-quality volume in less than 1 second. We then exclude empty spaces with a density threshold $\tau$ and randomly sample $M$ points as kernel positions. Following~\cite{kerbl20233d}, we set the scales of Gaussians as the nearest neighbor distances and assume no rotation. The central densities are queried from the FDK volume. We empirically scale down the queried densities with $k$ to compensate for the overlay between kernels.

\subsection{Training radiative Gaussians}
\label{sec: training radiative Gaussians}

\begin{figure}[t]
    \centering
    \includegraphics[width=\linewidth]{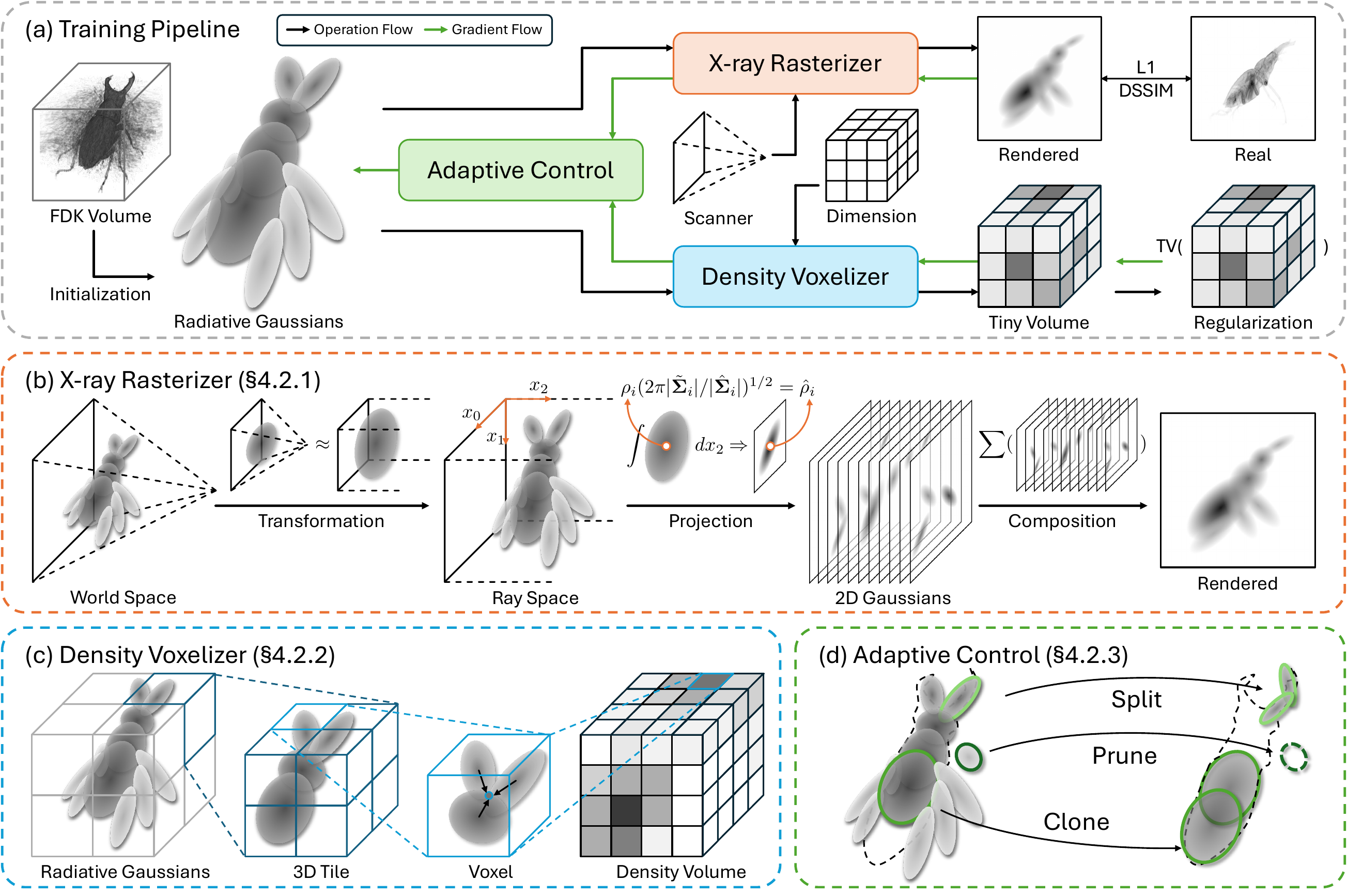}
    \caption{Training pipeline of R$^2$-Gaussian. (a) Overall training pipeline. (b) X-ray rasterization for projection rendering. (c) Density voxelization for volume retrieval. (d) Modified adaptive control.}
    \label{fig: pipeline}
\end{figure}

Our training pipeline is shown in~\cref{fig: pipeline}. Radiative Gaussians are first initialized from an FDK volume. We then rasterize projections for photometric losses and voxelize tiny density volumes for 3D regularization. Adaptive control is used to densify Gaussians for better representation. After training, we voxelize density volumes of the target size for evaluation.

\subsubsection{X-ray rasterization}
\label{sec: x-ray rasterization}

This section focuses on the theoretical derivation of X-ray rasterization $\mathcal{R}$. As discussed in~\cref{sec: x-ray imaging}, the pixel value of a projection is the integral of density along the corresponding ray path. We substitute~\cref{equ: density as sum of 3d Gaussians} into~\cref{equ: x-ray imaging}, yielding
\begin{equation}
\label{equ: sum of integration of 3d Gaussians}
    I_{r}(\mathbf{r}) = \int \sum_{i=1}^{M} G^3_{i}(\mathbf{r}(t)|\rho_i, \mathbf{p}_{i}, \mathbf{\Sigma}_{i}) d t  = \sum_{i=1}^{M} \int G^3_{i} (\mathbf{r}(t)|\rho_i, \mathbf{p}_{i}, \mathbf{\Sigma}_{i}) d t,
\end{equation}
where $I_{r}(\mathbf{r})$ is the rendered pixel value. This implies that we can individually integrate each 3D Gaussian to rasterize an X-ray projection. Note that $t_n$ and $t_f$ in~\cref{equ: x-ray imaging} are neglected because we assume all Gaussians are bounded inside the target space.

\paragraph{Transformation} 
Since a cone-beam X-ray scanner can be modeled similarly to a pinhole camera, we follow~\cite{zwicker2002ewa} to transfer Gaussians from the world space to the ray space. In ray space, the viewing rays are parallel to the third coordinate axis, facilitating analytical integration. Due to the non-Cartesian nature of ray space, we employ the local affine transformation to~\cref{equ: sum of integration of 3d Gaussians}, yielding
\begin{equation}
\label{equ: transfer from world space to ray space}
    I_{r}(\mathbf{r}) \approx \sum_{i=1}^{M} \int G^3_{i} (\tildebf{x}|\rho_i, \underbrace{ \phi(\mathbf{p})}_{\tildebf{p}_{i}}, \underbrace{\vphantom{\phi(\mathbf{p})} \mathbf{J}_{i}\mathbf{W}\mathbf{\Sigma}_{i}\mathbf{W}^{\top}\mathbf{J}_{i}^{\top}}_{\tildebf{\Sigma}_{i}}) d x_2,
\end{equation}
where $\smash{\tildebf{x}=[x_0,x_1,x_2]^\top}$ is a point in ray space, $\smash{\tildebf{p}_{i}\in \mathbb{R}^3}$ is the new Gaussian position obtained through projective mapping $\phi$, and $\smash{\tildebf{\Sigma}_{i}\in \mathbb{R}^{3\times3}}$ is the new Gaussian covariance controlled by local approximation matrix $\mathbf{J}_{i}$ and viewing transformation matrix $\mathbf{W}$. Refer to~\cref{sec: supp transform} for determining $\phi$, $\mathbf{J}_{i}$, and $\mathbf{W}$ from scanner parameters.

\paragraph{Projection and composition} A good property of \textit{normalized} 3D Gaussian distribution is that its integral along one coordinate axis yields a normalized 2D Gaussian distribution. Substitute~\cref{equ: one gaussian} into~\cref{equ: transfer from world space to ray space} and we have

\begin{equation}
\label{equ: sum of 2d gaussians}
\begin{aligned}
     I_{r}(\mathbf{r}) & \approx \sum_{i=1}^{M} \rho_{i} (2\pi)^{\frac{3}{2}}|\tildebf{\Sigma}_{i}|^{\frac{1}{2}}  \int  \underbrace{\frac{1}{(2\pi)^{\frac{3}{2}}|\tildebf{\Sigma}_{i}|^{\frac{1}{2}}} \exp \left( -\frac{1}{2}(\tildebf{x}-\tildebf{p}_{i})^\top \tildebf{\Sigma}^{-1}_{i}(\tildebf{x}-\tildebf{p}_{i})\right)}_{\text{Normalized 3D Gaussian distribution}} d x_2 \\
    & = \sum_{i=1}^{M} \rho_{i} (2\pi)^{\frac{3}{2}}|\tildebf{\Sigma}_{i}|^{\frac{1}{2}} \underbrace{\frac{1}{2\pi|\hatbf{\Sigma}_{i}|^{\frac{1}{2}}} \exp \left( -\frac{1}{2}(\hatbf{x}-\hatbf{p}_{i})^\top \hatbf{\Sigma}^{-1}_{i}(\hatbf{x}-\hatbf{p}_{i})\right)}_{\text{Normalized 2D Gaussian distribution}} \\
     & = \sum_{i=1}^{M} G^{2}_{i}(\hatbf{x}|\underbrace{\sqrt{\frac{2\pi|\tildebf{\Sigma}_{i}|}{|\hatbf{\Sigma}_{i}|}} \rho_{i}}_{\hat{\rho}_i}, \hatbf{p}_{i}, \hatbf{\Sigma}_{i}),
\end{aligned}
\end{equation}

where $\smash{\hatbf{x}\in \mathbb{R}^2}$, $\smash{\hatbf{p}\in \mathbb{R}^2}$, $\smash{\hatbf{\Sigma}\in \mathbb{R}^{2\times2}}$ are obtained by dropping the third rows and columns of their counterparts $\smash{\tildebf{x}}$, $\smash{\tildebf{p}}$, and $\smash{\tildebf{\Sigma}}$, respectively.~\cref{equ: sum of 2d gaussians} shows that an X-ray projection can be rendered by simply summing 2D Gaussians instead of alpha-compositing them in natural light imaging.

\begin{wrapfigure}{r}{0.25\textwidth}
    \centering
    \includegraphics[width=0.25\textwidth]{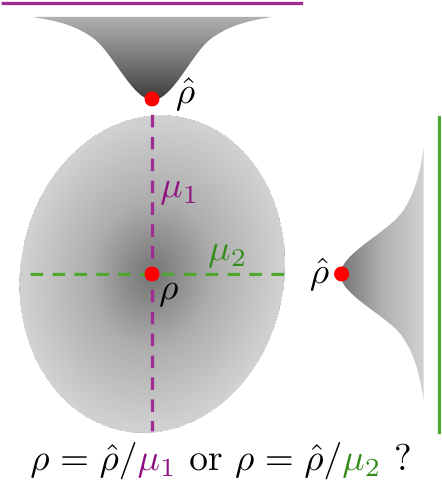}
    \caption{Density inconsistency in 3DGS.}
    \label{fig: inconsistency}
\end{wrapfigure}

\paragraph{Integration bias} 
During the projection, a key difference between our 2D Gaussian and the original one in 3DGS is the central density (opacity) $\smash{\hat{\rho}_{i}}$. As shown in~\cref{equ: sum of 2d gaussians}, we scale the density with a covariance-related factor $\smash{\mu_{i}=(2\pi|\tildebf{\Sigma}_{i}|/|\hatbf{\Sigma}_{i}|)^{1/2}}$: $\hat{\rho}_{i}=\mu_{i}\rho_{i}$, while 3DGS does not. This implies that 3DGS, in fact, learns an integrated density in the 2D image plane rather than the actual one in 3D space. This integration bias, though having a negligible impact on imaging rendering, leads to significant inconsistency in density retrieval. We demonstrate the inconsistency with a simplified 2D-to-1D projection in~\cref{fig: inconsistency}. When attempting to recover the central density $\rho$ in 3D space with $\rho_{i}=\hat{\rho}_{i}/\mu_{j}$, we find different views ($\mu_{j}$) lead to different results. This violates the isotropic nature of $\rho_{i}$, preventing us from determining the correct value. In contrast, our method assigns the actual 3D density to the kernel and forwardly computes the 2D projection, thus fundamentally solving the issue. While conceptually simple, implementing our idea requires substantial engineering efforts, including reprogramming all backpropagation routines in CUDA.

\subsubsection{Density voxelization}
\label{sec: density voxelization}

We develop a voxelizer $\mathcal{V}$ to efficiently query a density volume $\mathbf{V} \in \mathbb{R}^{X \times Y \times Z}$ from radiative Gaussians: $\mathbf{V} = \mathcal{V}(\mathbb{G}^3)$. Inspired by voxelizers used in RGB tasks~\cite{tang2024dreamgaussian}, our voxelizer first partitions the target space into multiple $8 \times 8 \times 8$ 3D tiles. It then culls Gaussians, retaining those with a $99\%$ confidence of intersecting the tile. In each 3D tile, voxel values are parallelly computed by summing the contributions of nearby kernels with~\cref{equ: density as sum of 3d Gaussians}. We implement the voxelizer and its backpropagation in CUDA, making it differentiable for optimization. This design not only accelerates the query process ($>100$ FPS) but also allows us to regularize radiative Gaussians with 3D priors.

\subsubsection{Optimization}
\label{sec: optimization}

We optimize radiative Gaussians using stochastic gradient descent. Besides photometric L1 loss $\mathcal{L}_{1}$ and D-SSIM loss $\mathcal{L}_{ssim}$~\cite{wang2004image}, 
we further incorporate a 3D total variation (TV) regularization~\cite{rudin1992nonlinear} $\mathcal{L}_{tv}$ as a homogeneity prior for tomography. At each training iteration, we randomly query a tiny density volume $\mathbf{V}_{tv} \in \mathbb{R}^{D \times D \times D}$ (same spacing as the target output) and minimize its total variation. The overall training loss is defined as:
\begin{equation}
\label{equ: loss}
\mathcal{L}_{total} = \mathcal{L}_{1}(\mathbf{I}_{r}, \mathbf{I}_{m}) + \lambda_{ssim} \mathcal{L}_{ssim}(\mathbf{I}_{r}, \mathbf{I}_{m}) + \lambda_{tv} \mathcal{L}_{tv}(\mathbf{V}_{tv}),
\end{equation}
where $\mathbf{I}_{r}$, $\mathbf{I}_{m}$, $\lambda_{ssim}$ and $\lambda_{tv}$ are rendered projection, measured projection, D-SSIM weight, and TV weight, respectively. Adaptive control is employed during training to enhance object representation. We remove empty Gaussians and densify (clone or split) those with large loss gradients. Considering objects such as human organs have extensive homogeneous areas, we do not prune large Gaussians. As for densification, we halve the densities of both original and replicated Gaussians. This strategy mitigates the sudden performance drop caused by new Gaussians and hence stabilizes training.
\section{Experiments}
\label{sec: experiments}

\begin{table}[t]
\centering
\caption{Quantitative results on sparse-view tomography. We colorize the \boxcolorf{best}, \boxcolors{second-best}, and \boxcolort{third-best} numbers.}
\label{tab: main result}
\resizebox{\columnwidth}{!}{%
\begin{tabular}{@{}cccccccccc@{}}
\toprule
& \multicolumn{3}{c}{75-view} & \multicolumn{3}{c}{50-view} & \multicolumn{3}{c}{25-view} \\ \cmidrule(lr){2-4} \cmidrule(lr){5-7} \cmidrule(lr){8-10}
\multirow{-2}{*}{Methods} & PSNR$\uparrow$ & SSIM$\uparrow$ & Time$\downarrow$ & PSNR$\uparrow$ & SSIM$\uparrow$ & Time$\downarrow$ & PSNR$\uparrow$ & SSIM$\uparrow$ & Time$\downarrow$ \\ \midrule
\multicolumn{10}{c}{Synthetic dataset} \\ \midrule
FDK~\cite{feldkamp1984practical} & 28.63 & 0.497 & - & 26.50 & 0.422 & - & 22.99 & 0.317 & - \\
SART~\cite{andersen1984simultaneous} & 36.06 & 0.897 & \cellcolor[HTML]{FFF6A9}4m41s & 34.37 & 0.875 & \cellcolor[HTML]{FFF6A9}3m36s & 31.14 & 0.825 & \cellcolor[HTML]{FFC991}1m47s \\
ASD-POCS~\cite{sidky2008image} & 36.64 & 0.940 & \cellcolor[HTML]{FF9396}2m25s & 34.34 & 0.914 & \cellcolor[HTML]{FF9396}1m52s & 30.48 & 0.847 & \cellcolor[HTML]{FF9396}56s \\
IntraTomo~\cite{zang2021intratomo} & 35.42 & 0.924 & 2h7m & 35.25 & 0.923 & 2h9m & \cellcolor[HTML]{FFF6A9}34.68 & \cellcolor[HTML]{FFF6A9}0.914 & 2h19m \\
NAF~\cite{zha2022naf} & 37.84 & 0.945 & 30m43s & 36.65 & 0.932 & 32m4s & 33.91 & 0.893 & 31m1s \\
SAX-NeRF~\cite{cai2023structure} & \cellcolor[HTML]{FFF6A9}38.07 & \cellcolor[HTML]{FFF6A9}0.950 & 13h5m & \cellcolor[HTML]{FFF6A9}36.86 & \cellcolor[HTML]{FFF6A9}0.938 & 13h5m & 34.33 & 0.905 & 13h3m \\
Ours (iter=10k) & \cellcolor[HTML]{FFC991}38.29 & \cellcolor[HTML]{FFC991}0.954 & \cellcolor[HTML]{FFC991}2m38s & \cellcolor[HTML]{FFC991}37.63 & \cellcolor[HTML]{FFC991}0.949 & \cellcolor[HTML]{FFC991}2m35s & \cellcolor[HTML]{FFC991}35.08 & \cellcolor[HTML]{FFC991}0.922 & \cellcolor[HTML]{FFF6A9}2m35s \\
Ours (iter=30k) & \cellcolor[HTML]{FF9396}38.88 & \cellcolor[HTML]{FF9396}0.959 & 8m21s & \cellcolor[HTML]{FF9396}37.98 & \cellcolor[HTML]{FF9396}0.952 & 8m14s & \cellcolor[HTML]{FF9396}35.19 & \cellcolor[HTML]{FF9396}0.923 & 8m28s \\ \midrule
\multicolumn{10}{c}{Real-world dataset} \\ \midrule
FDK~\cite{feldkamp1984practical} & 30.03 & 0.535 & - & 27.38 & 0.449 & - & 23.30 & 0.335 & - \\
SART~\cite{andersen1984simultaneous} & 34.42 & 0.845 & \cellcolor[HTML]{FFF6A9}5m11s & 33.61 & 0.827 & \cellcolor[HTML]{FFC991}3m28s & 31.52 & 0.790 & \cellcolor[HTML]{FFC991}1m47s \\
ASD-POCS~\cite{sidky2008image} & 36.33 & \cellcolor[HTML]{FFF6A9}0.868 & \cellcolor[HTML]{FF9396}2m43s & 34.58 & \cellcolor[HTML]{FFF6A9}0.861 & \cellcolor[HTML]{FF9396}1m49s & 31.32 & 0.810 & \cellcolor[HTML]{FF9396}56s \\
IntraTomo~\cite{zang2021intratomo} & 36.79 & 0.858 & 2h25m & \cellcolor[HTML]{FFF6A9}36.99 & 0.854 & 2h19m & \cellcolor[HTML]{FF9396}35.85 & \cellcolor[HTML]{FFC991}0.835 & 2h18m \\
NAF~\cite{zha2022naf} & \cellcolor[HTML]{FFC991}38.58 & 0.848 & 51m28s & 36.44 & 0.818 & 51m31s & 32.92 & 0.772 & 51m24s \\
SAX-NeRF~\cite{cai2023structure} & 34.93 & 0.854 & 13h21m & 34.89 & 0.840 & 13h23m & 33.49 & 0.793 & 13h25m \\
Ours (iter=10k) & \cellcolor[HTML]{FFF6A9}38.10 & \cellcolor[HTML]{FFC991}0.872 & \cellcolor[HTML]{FFC991}3m39s & \cellcolor[HTML]{FFC991}37.52 & \cellcolor[HTML]{FF9396}0.866 & \cellcolor[HTML]{FFF6A9}3m37s & \cellcolor[HTML]{FFC991}35.10 & \cellcolor[HTML]{FF9396}0.840 & \cellcolor[HTML]{FFF6A9}3m23s \\
Ours (iter=30k) & \cellcolor[HTML]{FF9396}39.40 & \cellcolor[HTML]{FF9396}0.875 & 14m16s & \cellcolor[HTML]{FF9396}38.24 & \cellcolor[HTML]{FFC991}0.864 & 13m52s & \cellcolor[HTML]{FFF6A9}34.83 & \cellcolor[HTML]{FFF6A9}0.833 & 12m56s \\ \bottomrule
\end{tabular}%
}
\end{table}

\subsection{Experimental settings}
\label{sec: expsetting}
\paragraph{Dataset} We conduct experiments on both synthetic and real-world datasets. For the synthetic dataset, we collect 15 real CT volumes, ranging from organisms to artificial objects. We then use the tomography toolbox TIGRE~\cite{biguri2016tigre} to synthesize X-ray projections and add Compton scatter and electric noise. For real-world experiments, we use three cases from the FIPS dataset~\cite{FIPS_CT_dataset}, each with 721 real projections. Since ground truth volumes are unavailable, we use FDK~\cite{feldkamp1984practical} to create pseudo-ground truth using all views and then subsample views for sparse-view experiments. We set 75, 50, and 25 views for both synthetic and real-world data as three sparse-view scenarios. Refer to~\cref{sec: supp details of dataset} for more details of datasets.

\paragraph{Implementation details} Our R$^2$-Gaussian is implemented in PyTorch~\cite{paszke2019pytorch} and CUDA~\cite{sanders2010cuda}, and trained with the Adam optimizer~\cite{KingBa15} for 30k iterations. Learning rates for position, density, scale, and rotation are initially set as 0.0002, 0.01, 0.005, and 0.001, respectively, and exponentially to 0.1 of their initial values. Loss weights are $\lambda_{ssim}=0.25$ and $\lambda_{tv}=0.05$. We initialize $M=50$k Gaussians with a density threshold $\tau=0.05$ and scaling term $k=0.15$. The TV volume size is $D=32$. Adaptive control runs from 500 to 15k iterations with a gradient threshold of 0.00005. All methods run on a single RTX3090 GPU. We evaluate reconstruction quality using PSNR and SSIM~\cite{wang2004image}, with PSNR calculated in 3D volume and SSIM averaged over 2D slices in axial, coronal, and sagittal directions. We also report the running time as a reflection of efficiency.

\subsection{Results and evaluation}

\begin{figure}[t]
    \centering
    \includegraphics[width=\linewidth]{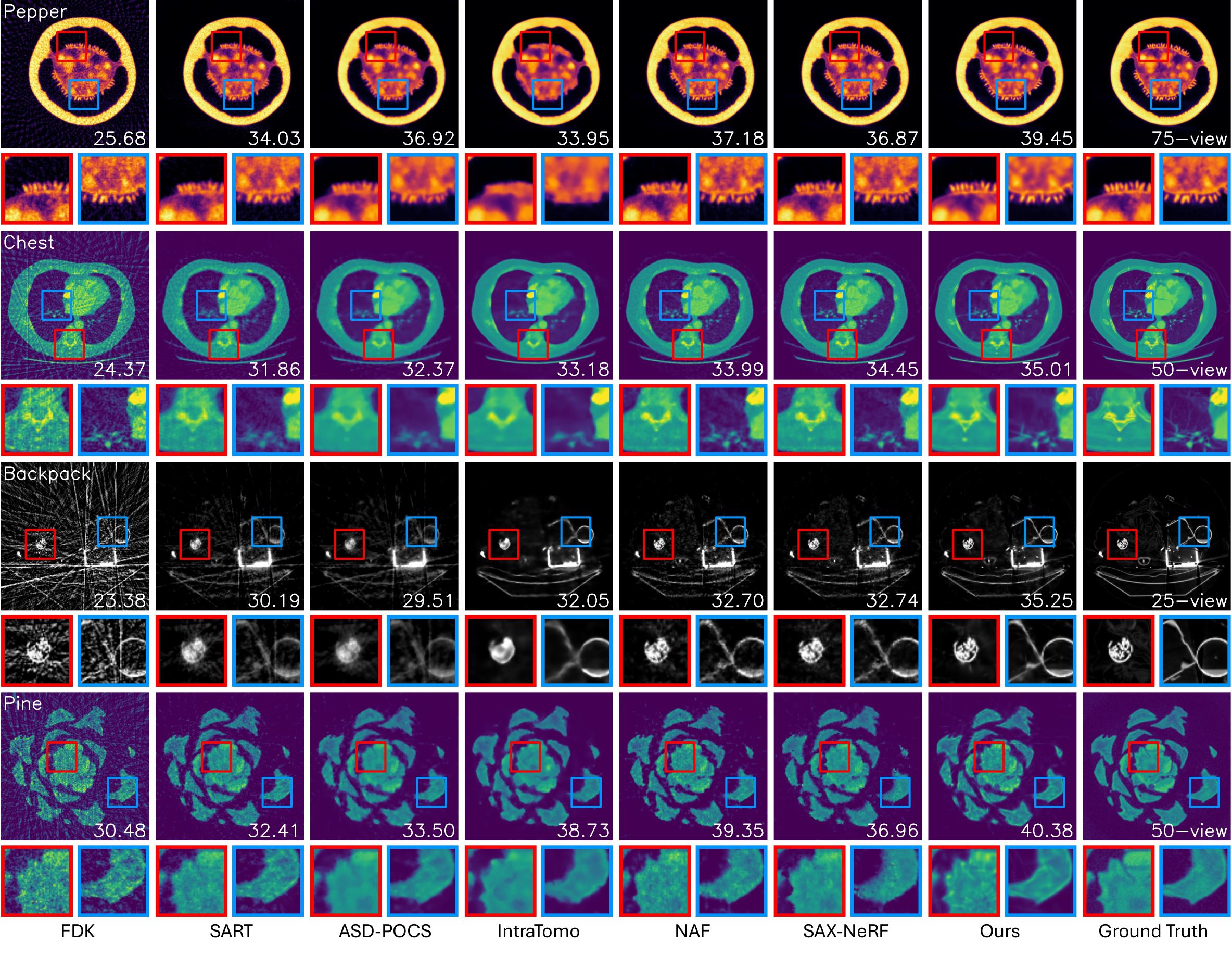}
    \caption{Colorized slice examples of different methods with PSNR (dB) shown at the bottom right of each image. The first three rows are from the synthetic dataset and the last row is from the real-world dataset. Our method recovers more details and suppresses artifacts.
    }
    \label{fig: qualitative}
\end{figure}

For fairness, we do not compare methods that require external training data but focus on those that solely use 2D projections of arbitrary objects. We compare R$^2$-Gaussian with three traditional methods (FDK~\cite{feldkamp1984practical}, SART~\cite{andersen1984simultaneous}, ASD-POCS~\cite{sidky2008image}) and three SOTA NeRF-based methods (IntraTomo~\cite{zang2021intratomo}, NAF~\cite{zha2022naf}, SAX-NeRF~\cite{cai2023structure}). \cref{tab: main result} reports the quantitative results on sparse-view tomography. Note that we do not report the running time for FDK as it is instant. R$^2$-Gaussian achieves the best performance across all synthetic and most real-world experiments. Specifically, our method delivers a 0.93 dB higher PSNR than SAX-NeRF, on the synthetic dataset, and a 0.95 dB improvement over IntraTomo on the real-world dataset. It is also worth noting that our 50-view results are already on par with the 75-view results of other methods. Regarding efficiency, our method converges to optimal results in 15 minutes, which is 3.7$\times$ faster than the most efficient NeRF-based method, NAF. Surprisingly, it takes less than 4 minutes to surpass other methods, which is even faster than the traditional algorithm SART. \cref{fig: qualitative} shows the visual comparisons of different methods. FDK and SART introduce streak artifacts, while ASD-POCS and IntraTomo blur structural details. NAF and SAX-NeRF are better than other baseline methods but have salt-and-pepper noise. In comparison, our method successfully recovers sharp details, e.g., ovules of pepper, and maintains good smoothness for homogeneous areas, e.g., muscles in the chest.

\subsection{Ablation study}
\label{sec: ablation study}

\begin{figure}[b]
    \centering
    \includegraphics[width=\linewidth]{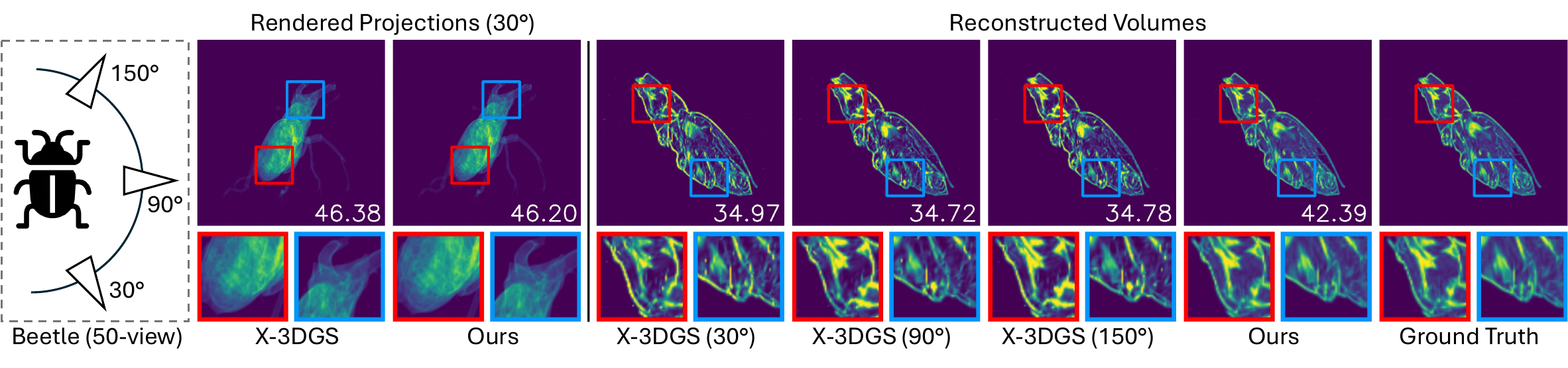}
    \caption{Results of X-3DGS and our method with PSNR (dB) indicated on each image. We show slices of X-3DGS queried from three viewing angles. Although X-3DGS can produce plausible X-ray projections, its reconstructed volume lacks view consistency and exhibits poor quality.}
    \label{fig: ablation_integration}
\end{figure}

\paragraph{Integration bias} To demonstrate the impact of integration bias discussed in \cref{sec: x-ray rasterization}, we develop an X-ray version of 3DGS (X-3DGS) that uses X-ray rendering while retaining the biased 3D-to-2D Gaussian projection. We use the same voxelizer in~\cref{sec: density voxelization} to extract volumes. Before voxelization, we divide the learned density of each Gaussian by the mean scaling factor $\mu$ of all training views. \cref{tab:x-3dgs vs ours} shows that rectifying integration bias benefits both 2D rendering (+3.15 dB PSNR) and 3D reconstruction (+17.77 dB PSNR). \cref{fig: ablation_integration} visualize rendering and reconstruction results. While X-3DGS renders reasonable 2D projections, its reconstruction quality is significantly worse than ours. Besides, there are notable discrepancies in slices queried from different views. The conflicting 2D and 3D performances indicate that X-3DGS, despite fitting images well, does not accurately model the density field.  In contrast, our method learns the actual view-independent density, eliminating inconsistencies and ensuring unbiased object representation.

\begin{table}[]
\centering
\caption{Quantitative results of X-3DGS and our method on the synthetic dataset.}
\label{tab:x-3dgs vs ours}
\begin{tabular}{@{}ccccccc@{}}
\toprule
\multirow{2}{*}{} & \multicolumn{2}{c}{75-view} & \multicolumn{2}{c}{50-view} & \multicolumn{2}{c}{25-view} \\ \cmidrule(lr){2-3} \cmidrule(lr){4-5} \cmidrule(l){6-7} 
 & X-3DGS & Ours & X-3DGS & Ours & X-3DGS & Ours \\ \midrule
2D PSNR$\uparrow$ & 49.97 & \cellcolor[HTML]{FF9396}50.54 & 47.26 & \cellcolor[HTML]{FF9396}49.70 & 39.84 & \cellcolor[HTML]{FF9396}46.28 \\
2D SSIM$\uparrow$ & \cellcolor[HTML]{FF9396}0.987 & 0.986 & 0.984 & \cellcolor[HTML]{FF9396}0.986 & 0.967 & \cellcolor[HTML]{FF9396}0.982 \\
3D PSNR$\uparrow$ & 23.40 & \cellcolor[HTML]{FF9396}38.86 & 21.24 & \cellcolor[HTML]{FF9396}37.98 & 14.07 & \cellcolor[HTML]{FF9396}35.17 \\
3D SSIM$\uparrow$ & 0.660 & \cellcolor[HTML]{FF9396}0.959 & 0.562 &\cellcolor[HTML]{FF9396}0.952 & 0.408 & \cellcolor[HTML]{FF9396}0.923 \\ \bottomrule
\end{tabular}%
\end{table}

\paragraph{Component analysis} We conduct ablation experiments to assess the effect of FDK initialization (Init.), modified adaptive control (AC), and total variation regularization (Reg.) on performance. The baseline model excludes these components and uses randomly generated Gaussians for initialization. Experiments are performed under the 50-view condition, evaluating PSNR, SSIM, training time, and Gaussian count (Gau.). Results are listed in~\cref{tab: ablation}. FDK initialization boosts PSNR by $0.9$ dB. Adaptive control improves quality but prolongs training due to more Gaussians. TV regularization increases SSIM by reducing artifacts and promoting smoothness. Overall, our full model outperforms the baseline, improving PSNR by 1.51 dB and SSIM by 0.018, with training time under 9 minutes.

\begin{table}[t]
    \centering
    \begin{minipage}{0.5\textwidth}
        \centering
        \caption{Ablation results with our choices in bold.}
        \footnotesize
        \begin{tabular}{@{}ccccc@{}}
            \toprule
            Synthetic & PSNR$\uparrow$ & SSIM$\uparrow$ & Time$\downarrow$ & Gau. \\ \midrule
            Baseline (B) & 36.47 & 0.934 & \cellcolor[HTML]{FF9396}4m57s & 50k \\
            B+Init. & \cellcolor[HTML]{FFC991}37.37 & \cellcolor[HTML]{FFC991}0.944 & \cellcolor[HTML]{FFC991}5m29s & 50k \\
            B+AC & \cellcolor[HTML]{FFF6A9}37.33 & 0.942 & 7m33s & 70k \\
            B+Reg. & 36.79 & \cellcolor[HTML]{FFF6A9}0.943 & \cellcolor[HTML]{FFF6A9}6m30s & 50k \\
            \textbf{Full model} & \cellcolor[HTML]{FF9396}37.98 & \cellcolor[HTML]{FF9396}0.952 & 8m37s & 68k \\ \midrule
            $M$=5k & 37.44 & 0.946 & 9m18s & 32k \\
            $M$=10k & 37.56 & 0.948 & \cellcolor[HTML]{FFC991}8m59s & 35k \\
            $\boldsymbol{M}\textbf{=50k}$ & \cellcolor[HTML]{FFC991}37.98 & \cellcolor[HTML]{FFC991}0.952 & \cellcolor[HTML]{FF9396}8m14s & 68k \\
            $M$=100k & \cellcolor[HTML]{FF9396}38.03 & \cellcolor[HTML]{FF9396}0.953 & \cellcolor[HTML]{FFF6A9}9m4s & 112k \\
            $M$=200k & \cellcolor[HTML]{FFF6A9}37.82 & \cellcolor[HTML]{FFF6A9}0.949 & 9m54s & 206k \\ \midrule
            $\lambda_{tv}$=0 & 37.66 & 0.948 & \cellcolor[HTML]{FF9396}7m9s & 68k \\
            $\lambda_{tv}$=0.01 & \cellcolor[HTML]{FFC991}37.88 & \cellcolor[HTML]{FFF6A9}0.950 & 8m21s & 68k \\
            $\boldsymbol{\lambda_{tv}}$\textbf{=0.05} & \cellcolor[HTML]{FF9396}37.98 & \cellcolor[HTML]{FF9396}0.952 & \cellcolor[HTML]{FFF6A9}8m14s & 68k \\
            $\lambda_{tv}$=0.1 & \cellcolor[HTML]{FFF6A9}37.73 & \cellcolor[HTML]{FFC991}0.951 & \cellcolor[HTML]{FFC991}8m11s & 68k \\
            $\lambda_{tv}$=0.15 & 37.40 & 0.949 & 8m27s & 69k \\ \midrule
            $D$=8 & 37.74 & 0.949 & \cellcolor[HTML]{FF9396}7m56s & 68k \\
            $D$=16 & \cellcolor[HTML]{FFC991}37.94 & \cellcolor[HTML]{FFF6A9}0.950 & \cellcolor[HTML]{FFF6A9}8m18s & 68k \\
            $\boldsymbol{D}$\textbf{=32} & \cellcolor[HTML]{FF9396}37.98 & \cellcolor[HTML]{FF9396}0.952 & \cellcolor[HTML]{FFC991}8m14s & 68k \\
            $D$=48 & \cellcolor[HTML]{FFF6A9}37.90 & \cellcolor[HTML]{FFC991}0.951 & 9m34s & 67k \\
            $D$=64 & 37.82 & 0.949 & 11m35s & 67k \\ \bottomrule
        \end{tabular}%
        \label{tab: ablation}
    \end{minipage}
    \hfill
    \begin{minipage}{0.48\textwidth}
        \includegraphics[width=\linewidth]{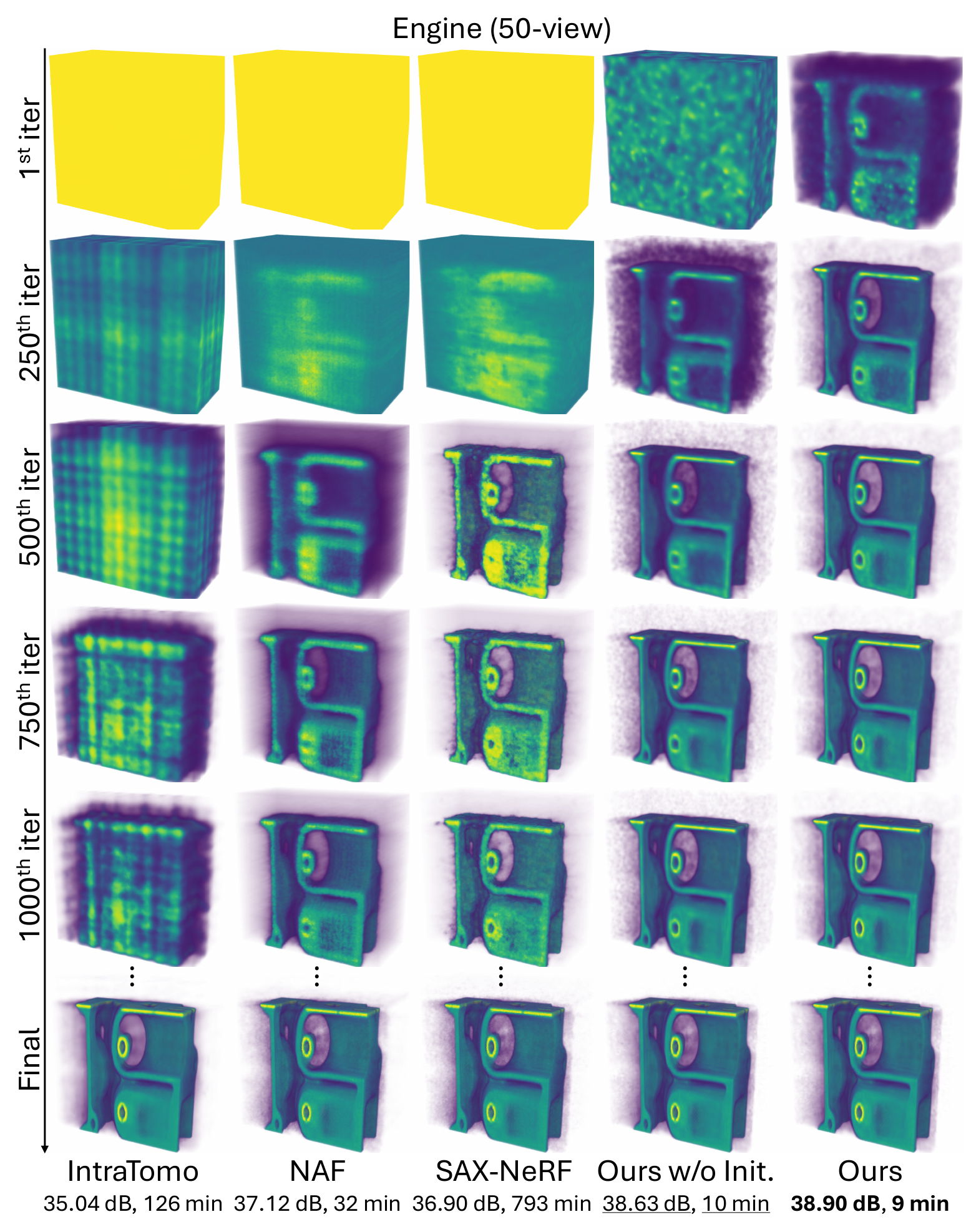}
        \captionof{figure}{Results of NeRF-based methods and our R$^2$-Gaussian at different iterations.}
        \label{fig: convergence}
    \end{minipage}
\end{table}

\paragraph{Parameter analysis} We perform parameter analysis on the number of initialized Gaussians $M$, TV loss weight $\lambda_{tv}$, and TV volume size $D$. The results are shown in the last three blocks of~\cref{tab: ablation}. R$^2$-Gaussian achieves good quality-efficiency balance at 50k initialized Gaussians. A TV loss weight of $\lambda_{tv}=0.05$ improves reconstruction, but larger values can lead to degradation. The training time increases with TV volume size while the performance peaks at $D=32$.

\paragraph{Convergence analysis} \cref{fig: convergence} compares the results of NeRF-based methods and our R$^2$-Gaussian method at different iterations. Our method, both with and without FDK initialization, converges significantly faster, displaying sharp details by the 500th iteration when other methods still exhibit artifacts and blurriness. Notably, the FDK initialization offers a rough structure before training, which further accelerates convergence and enhances reconstruction quality. Finally, our method outperforms others in both performance and efficiency, achieving the highest PSNR of 38.90 dB in 9 minutes.

\section{Discussion and conclusion}
\label{sec: limitation and conclusion}

\paragraph{Discussion} R$^2$-Gaussian inherits some limitations from 3DGS, such as varying training time across modalities, needle-like artifacts under extremely sparse-view conditions, and suboptimal extrapolation for other tomography tasks. Besides, we have not considered calibration errors regarding the scanned geometry and anisotropic physical effects such as Compton scattering. More details are discussed in~\cref{sec: details of limitation}. Despite these limitations, our method's superior performance and fast speed make it valuable for real-world applications for medical diagnosis and industrial inspection.

\paragraph{Conclusion} This paper presents R$^2$-Gaussian, a novel 3DGS-based framework for sparse-view tomographic reconstruction. We identify and rectify a previously overlooked integration bias of standard 3DGS, which hinders accurate density retrieval. Furthermore, we enhance 3DGS for tomography by introducing new kernels, devising X-ray rasterization functions, and developing a differentiable voxelizer. Our R$^2$-Gaussian surpasses state-of-the-art methods in both reconstruction quality and training speed, demonstrating its potential for real-world applications. Crucially, we speculate that the newly found integration bias may be pervasive across all 3DGS-related research. Consequently, our rectification technique could benefit more tasks beyond computed tomography.

\section*{Acknowledgments}
The research is funded in part by ARC Discovery Grant (grant ID: DP220100800) of the Australia Research Council. 

\bibliographystyle{plainnat}
\bibliography{mybib}

\clearpage

\appendix

\section{Transformation module in X-ray rasterization}
\label{sec: supp transform}
\begin{figure}[h]
    \centering
    \includegraphics[width=0.6\linewidth]{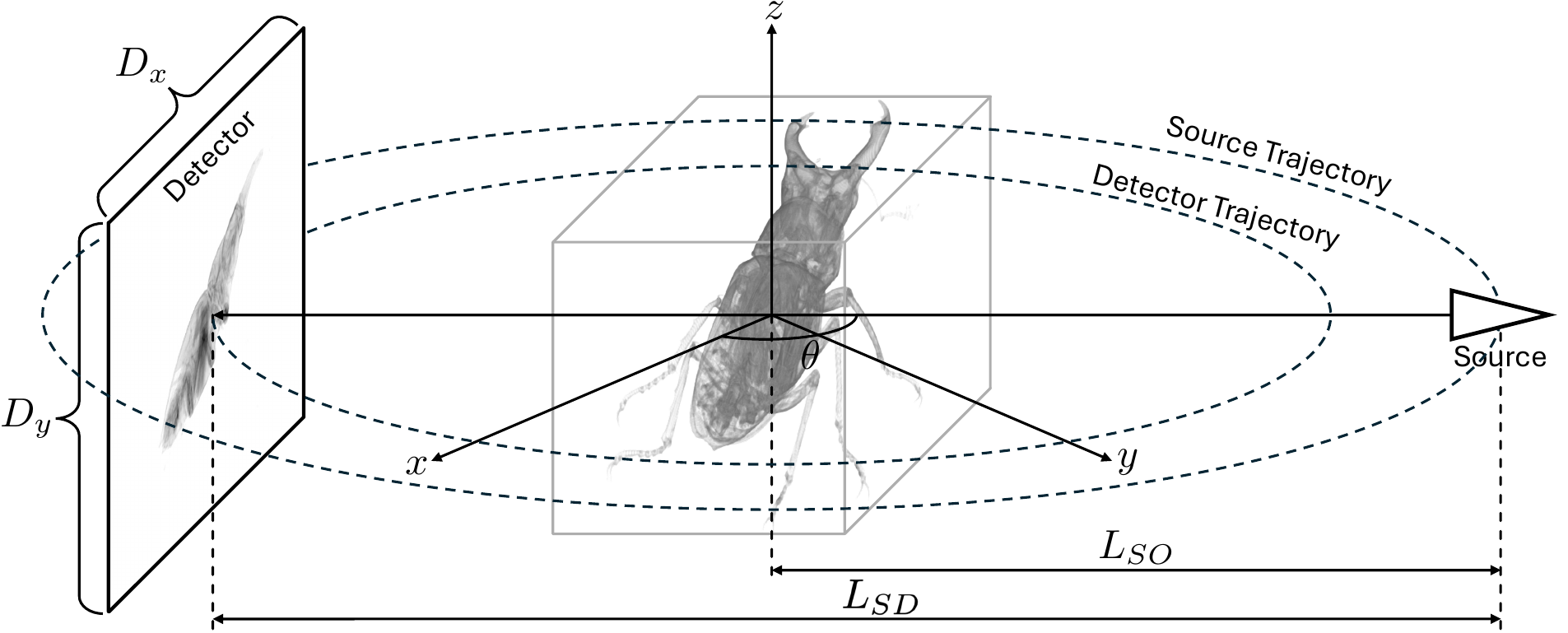}
    \caption{Configuration of a cone-beam CT scanner.}
    \label{fig: scanner}
\end{figure}

The configuration of a cone beam CT scanner is shown in~\cref{fig: scanner}. The X-ray source and detector plane rotate around the z-axis, resembling a pinhole camera model. Therefore, we can formulate the field-of-view (FOV) of a scanner as
\begin{equation}
    FOV_{x}=2\cdot \arctan(\frac{D_{x}}{2L_{SD}}), FOY_{y}=2\cdot \arctan(\frac{D_{y}}{2L_{SD}}).
\end{equation}
Here, $(D_{x}, D_{y})$ is the physical size of the detector plane, and $L_{SD}$ is the distance between the source and the detector. Following~\cite{kerbl20233d}, we then use FOVs to determine the projection mapping $\phi$.

To get Gaussians in the ray space, we first transfer them from the world space to the scanner space. The scanner space is defined such that its origin is the X-ray source, and its z-axis points to the projection center. The transformation matrix $\mathbf{T}$ from the world space to the scanner space is

\begin{equation}
\mathbf{T}=\begin{bmatrix}
\mathbf{W} &  \mathbf{t} \\ 
0 & 1 \\
\end{bmatrix}, 
\mathbf{W}=\begin{bmatrix}
 -\sin\theta& \cos\theta &  0\\ 
0 & 0 & -1\\ 
-\cos \theta & -\sin \theta & 0
\end{bmatrix},
\mathbf{t}=\begin{bmatrix}
0\\ 
0\\ 
 L_{SO}
\end{bmatrix}.
\end{equation}
Here, $\phi$ is the rotation angle, and $L_{SO}$ is the distance between the source and the object. Next, we apply local approximation on each Gaussian. The Jacobian of the affine approximation $\mathbf{J}_{i}$ is the same as Eq. (29) in~\cite{zwicker2002ewa}. Finally, we have the Gaussian in the ray space with new position $\tildebf{p}$ and covariance $\tildebf{\Sigma}_{i}$ as
\begin{equation}
     \tildebf{p}_{i} = \phi(\mathbf{p}), \tildebf{\Sigma}_{i}=\mathbf{J}_{i}\mathbf{W}\mathbf{\Sigma}_{i}\mathbf{W}^{\top}\mathbf{J}_{i}^{\top}.
\end{equation}

\section{Details of dataset}
\label{sec: supp details of dataset}

\begin{figure}[t]
    \centering
    \includegraphics[width=1.0\linewidth]{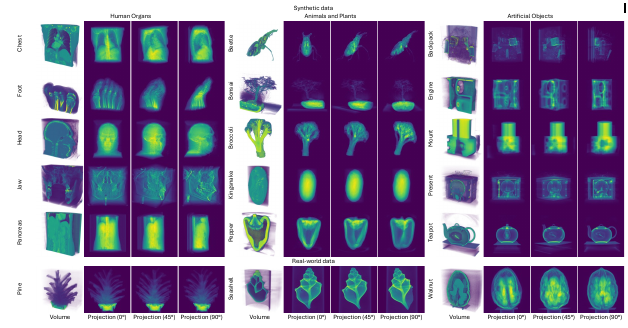}
    \caption{
    Datasets used for experiments. We show half volume and projection examples for each case.
    }
    \label{fig: gt demo}
\end{figure}

\paragraph{Synthetic data} We evaluate methods with various modalities, covering major CT applications such as medical diagnosis, biological research, and industrial inspection. The synthetic dataset consists of 15 cases across three categories: human organs (chest, foot, head, jaw, and pancreas), animals and plants (beetle, bonsai, broccoli, kingsnake, and pepper), and artificial objects (backpack, engine, present, teapot, and mount). The chest and pancreas scans are from LIDC-IDRI~\cite{armato2011lung} and Pancreas-CT~\cite{pancreasct}, respectively. Broccoli and pepper are obtained from X-Plant~\cite{verboven2022x}, and the rest are from SciVis~\cite{scivisdata}. Following~\cite{zha2022naf,cai2023structure}, we preprocess raw data by normalizing densities to $[0, 1]$ and resizing volumes to $256\times 256\times 256$. We then use the tomography toolbox TIGRE~\cite{biguri2016tigre} to capture $512\times 512$ projections in the range of $0^{\circ}\sim 360^{\circ}$. We add two types of noise: Gaussian (mean 0, standard deviation 10) as electronic noise of the detector and Poisson (lambda 1e5) as photon scattering noise. All volumes and their projection examples are shown in~\cref{fig: gt demo}.

\paragraph{Real-world data} We use FIPS~\cite{FIPS_CT_dataset}, a public dataset providing real 2D X-ray projections. FIPS includes three objects (pine~\cite{pine_data}, seashell~\cite{seashell_data}, and walnut~\cite{walnut_data}). Each case has 721 projections in the range of 0$^{\circ}$-360$^{\circ}$. We preprocess 2D projections by resizing them to $560x560$ and normalizing them to $[0,1]$. Since ground truth volumes are unavailable, we use FDK to create pseudo-ground truth with all views and then subsample 75/50/25 views for sparse-view experiments. The size of the target volume is $256\times 256\times 256$. 

\section{Implementation details of baseline methods}
\label{sec: more details about implmentation}
For fairness, we do not compare methods that require external training data but focus on those that solely use 2D projections of arbitrary objects. We run traditional algorithms FDK, SART, and ASD-POCS with GPU-accelerated tomographic toolbox TIGRE~\cite{biguri2016tigre}, and select three SOTA NeRF-based tomography methods. IntraTomo models the density field with a large MLP. NAF accelerates the training process by hash encoding. SAX-NeRF achieves plausible results with a line segment-based transformer. We use the official code of NAF and SAX-NeRF and conduct experiments with default hyperparameters. The IntraTomo implementation is sourced from the NAF repository. The training iterations of NeRF-based methods are set to 150k (default of NAF and SAX-NeRF). All methods are run on a single RTX 3090 GPU.

\section{More qualitative results}

\begin{figure}[h]
    \centering
    \includegraphics[width=1.0\linewidth]{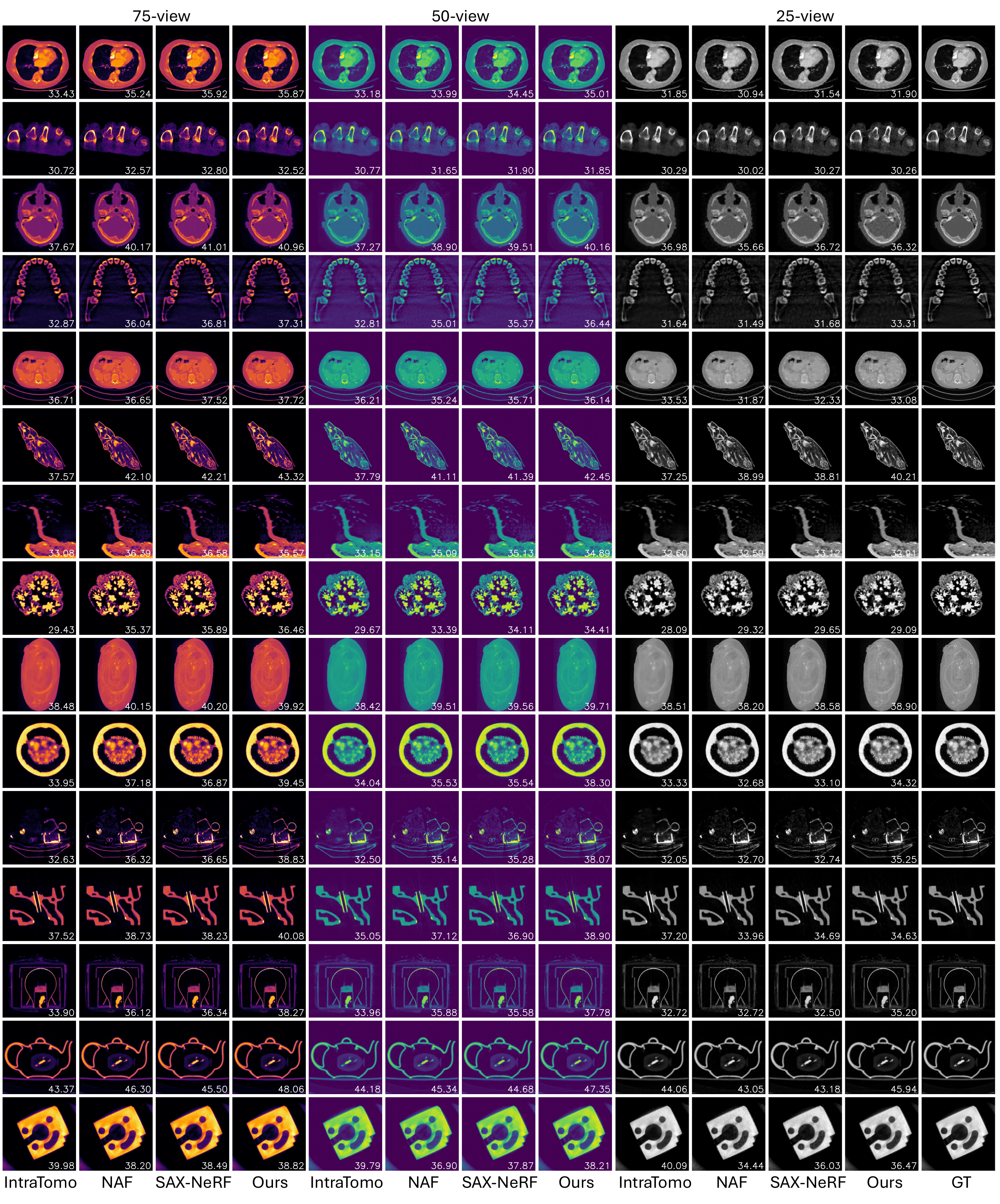}
    \caption{Reconstruction results of NeRF-based methods and our method on the synthetic dataset.}
    \label{fig: more_qualitative}
\end{figure}

\begin{figure}[h]
    \centering
    \includegraphics[width=1.0\linewidth]{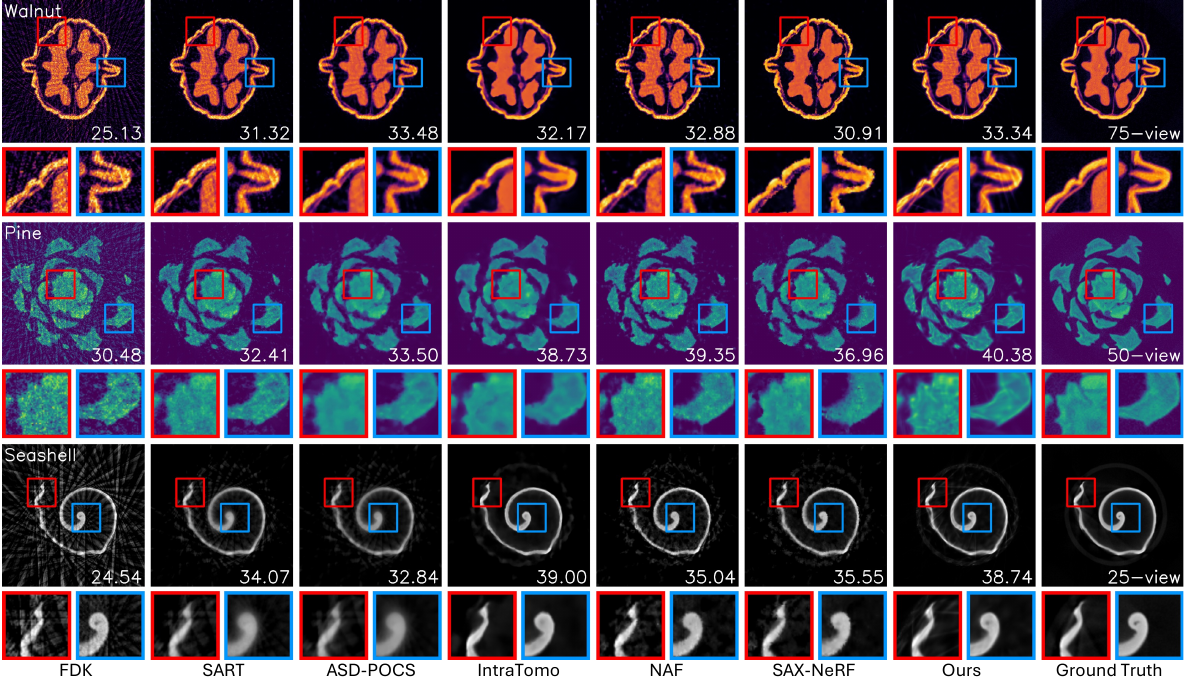}
    \caption{Reconstruction results on the real-world dataset.}
    \label{fig: more_qualitative_fips}
\end{figure}

\paragraph{Main results} We visualize more reconstruction results in~\cref{fig: more_qualitative} and~\cref{fig: more_qualitative_fips}. FDK and SART introduce notable streak artifacts, while ASD-POCS and IntraTomo blur structural details. NeRF-based solutions perform better than traditional methods but exhibit salt-and-pepper noise. In comparison, our method successfully recovers sharp details and maintains smoothness in homogeneous areas.

\begin{figure}[h]
    \centering
    \includegraphics[width=1.0\linewidth]{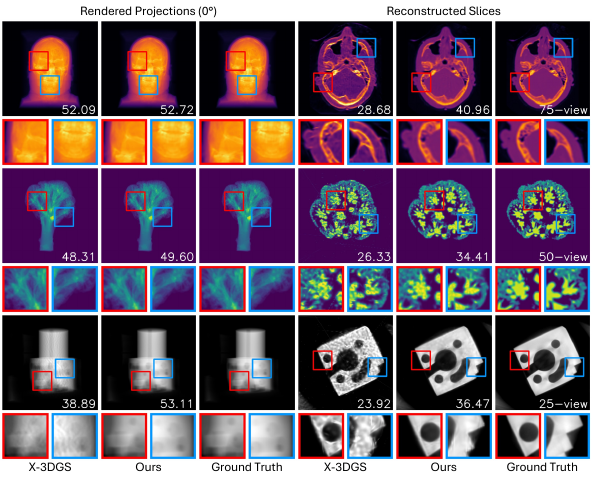}
    \caption{Qualitative comparison of X-3DGS and our method.}
    \label{fig: more_qualitative_x3dgs}
\end{figure}

\paragraph{Integration bias} We show more qualitative comparisons of X-3DGS and ours in~\cref{fig: more_qualitative_x3dgs}. Our method outperforms X-3DGS in both 2D rendering and 3D reconstruction.

\begin{figure}[h]
    \centering
    \includegraphics[width=0.85\linewidth]{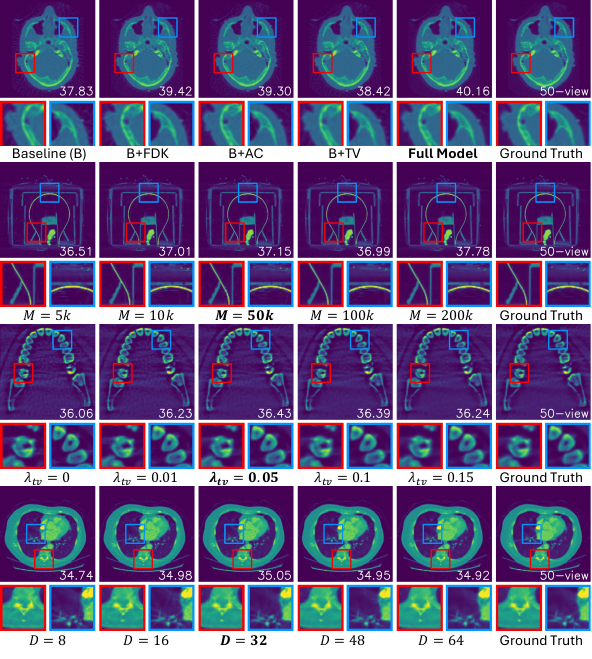}
    \caption{Quantitative comparison of different components and parameters.}
    \label{fig: more_ablation}
\end{figure}

\paragraph{Components and parameters} We visually compare different components and parameters in~\cref{fig: more_ablation}. Our newly introduced components improve the reconstruction quality. Our parameter setting also yields the best performance.

\begin{figure}[h]
    \centering
    \includegraphics[width=1.0\linewidth]{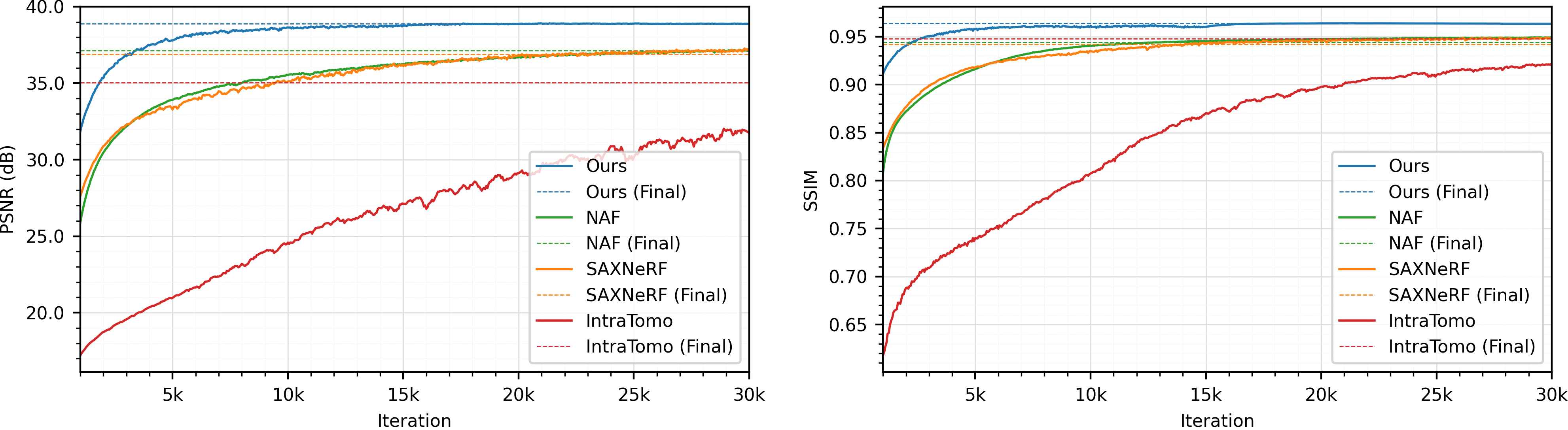}
    \caption{PSNR-iteration and SSIM-iteration plots of case $\textit{engine, 50-view}$.}
    \label{fig: psnr vs iter}
\end{figure}

\paragraph{Convergence analysis} We show the PSNR and SSIM plots in~\cref{fig: psnr vs iter}.  Our method converges significantly faster than NeRF-based methods and outperforms them in only 3000 iterations.

\section{More discussion of limitation}
\label{sec: details of limitation}

\begin{figure}[h]
    \centering
    \includegraphics[width=1.0\linewidth]{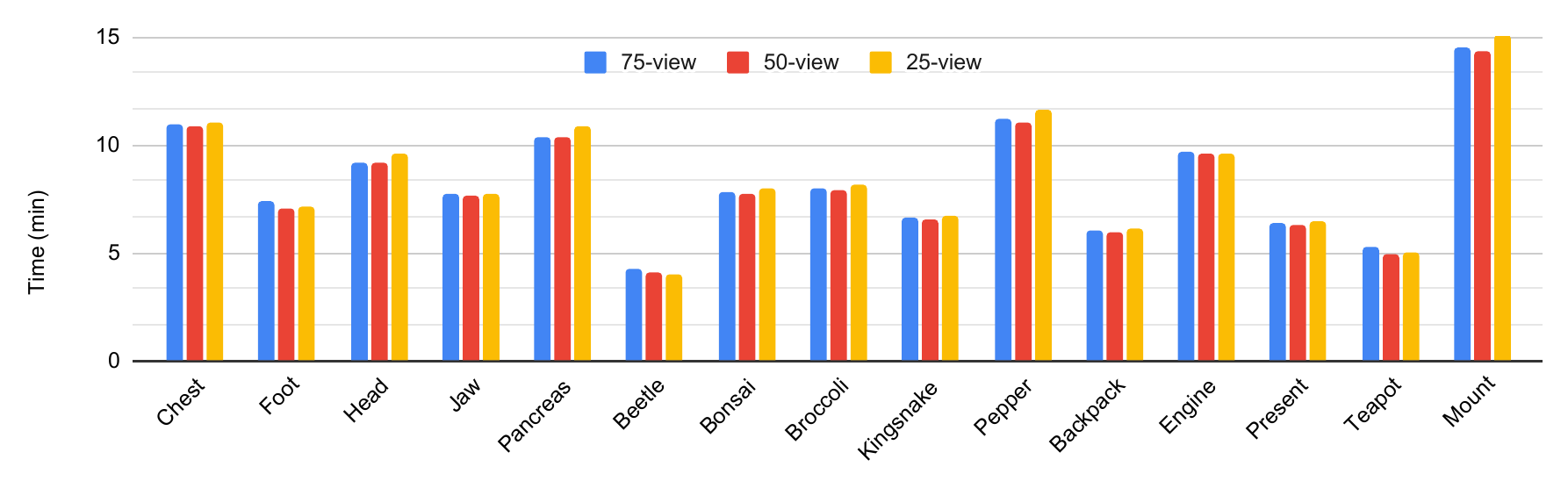}
    \caption{Training time on the synthetic dataset.}
    \label{fig: time all}
\end{figure}

\paragraph{Varying time} We present the training times for all cases in~\cref{fig: time all}. The training time varies across cases, primarily due to the different numbers of kernels used. Our method takes more time on objects with large homogeneous areas, such as the chest, pancreas, and mount, and less time on those with sparse structures, such as the beetle, backpack, and present.

\begin{figure}[h]
    \centering
    \includegraphics[width=1.0\linewidth]{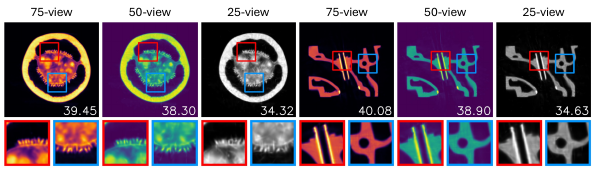}
    \caption{3DGS-based methods tend to introduce needle-like artifacts when there are insufficient amounts of images. PNSR (dB) is shown at the bottom right of each image.}
    \label{fig: artifact}
\end{figure}

\paragraph{Needle-like artifacts} 
While our method achieves the highest reconstruction quality, it introduces needle-like artifacts, especially under the 25-view condition (\cref{fig: artifact}). This suggests that some Gaussians may overfit specific X-rays. Similar artifacts are also observed in 3DGS~\cite{zhang2024pixel}.

\begin{figure}[h]
    \centering
    \includegraphics[width=1.0\linewidth]{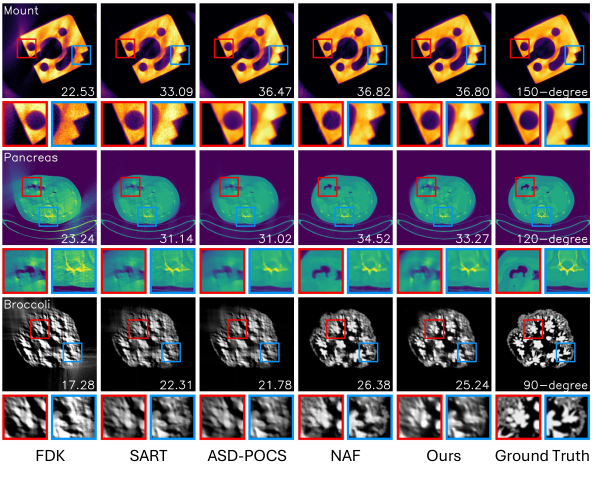}
    \caption{Visualization of reconstruction results under limited-angle scenarios. PNSR (dB) is shown at the bottom right of each image.}
    \label{fig: lact}
\end{figure}

\paragraph{Extrapolation ability} While this paper focuses on sparse-view CT (SVCT), we also test R$^2$-Gaussian on limited-angle CT (LACT), where the scanning range is constrained to less than $180^{\circ}$. Unlike SVCT, which highlights the interpolation ability of methods, LACT challenges their extrapolation ability, i.e., estimating unseen areas outside the scanning range. We generate 100 projections within ranges of $0^{\circ}\sim 150^{\circ}$, $0^{\circ}\sim 120^{\circ}$, and $0^{\circ}\sim 90^{\circ}$. The quantitative results in~\cref{tab: lact} show that our method has slightly lower PSNR but higher SSIM than the NeRF-based method NAF. Visualization results in~\cref{fig: lact} indicate that our method recovers more details in scanned areas but exhibits blurred artifacts in unseen areas. We attribute this performance drop to the nature of the networks and kernels. Given the gradient of a ray, NeRF updates the entire network while 3DGS individually optimizes intersected kernels. Thus, NeRF has better global awareness and consistency, while 3DGS is more local-oriented and has suboptimal extrapolation ability.

\begin{table}[b]
\centering
\caption{Quantitative evaluation on limited-angle tomography. We colorize the \boxcolorf{best}, \boxcolors{second-best}, and \boxcolort{third-best} numbers.}
\label{tab: lact}
\resizebox{\columnwidth}{!}{%
\begin{tabular}{@{}cccccccccc@{}}
\toprule
 & \multicolumn{3}{c}{$0^{\circ}\sim 150^{\circ}$} & \multicolumn{3}{c}{$0^{\circ}\sim 120^{\circ}$} & \multicolumn{3}{c}{$0^{\circ}\sim 90^{\circ}$} \\ \cmidrule(lr){2-4} \cmidrule(lr){5-7} \cmidrule(lr){8-10}
\multirow{-2}{*}{Methods} & PSNR$\uparrow$ & SSIM$\uparrow$ & Time$\downarrow$ & PSNR$\uparrow$ & SSIM$\uparrow$ & Time$\downarrow$ & PSNR$\uparrow$ & SSIM$\uparrow$ & Time$\downarrow$ \\ \midrule
FDK~\cite{feldkamp1984practical} & 26.83 & 0.570 & - & 24.00 & 0.566 & - & 21.22 & 0.547 & - \\
SART~\cite{andersen1984simultaneous} & \cellcolor[HTML]{FFF6A9}33.34 & 0.883 & \cellcolor[HTML]{FFC991}7m9s & \cellcolor[HTML]{FFF6A9}30.21 & 0.847 & \cellcolor[HTML]{FFC991}7m8s & \cellcolor[HTML]{FFF6A9}26.71 & 0.795 & \cellcolor[HTML]{FFC991}7m59s \\
ASD-POCS~\cite{sidky2008image} & 33.16 & \cellcolor[HTML]{FFF6A9}0.913 & \cellcolor[HTML]{FF9396}3m41s & 29.76 & \cellcolor[HTML]{FFF6A9}0.875 & \cellcolor[HTML]{FF9396}3m39s & 26.34 & \cellcolor[HTML]{FFF6A9}0.812 & \cellcolor[HTML]{FF9396}4m8s \\
NAF~\cite{zha2022naf} & \cellcolor[HTML]{FF9396}36.29 & \cellcolor[HTML]{FFC991}0.940 & 27m18s & \cellcolor[HTML]{FF9396}33.35 & \cellcolor[HTML]{FFC991}0.922 & 27m6s & \cellcolor[HTML]{FF9396}29.89 & \cellcolor[HTML]{FFC991}0.884 & 27m25s \\
Ours & \cellcolor[HTML]{FFC991}36.12 & \cellcolor[HTML]{FF9396}0.948 & \cellcolor[HTML]{FFF6A9}9m3s & \cellcolor[HTML]{FFC991}32.68 & \cellcolor[HTML]{FF9396}0.923 & \cellcolor[HTML]{FFF6A9}8m36s & \cellcolor[HTML]{FFC991}29.21 & \cellcolor[HTML]{FF9396}0.886 & \cellcolor[HTML]{FFF6A9}8m28s \\ \bottomrule
\end{tabular}%
}
\end{table}

\paragraph{Calibration error} 
In real-world applications, calibration errors can affect reconstruction quality. For example, tomography requires a reference image $I_{0}$ in \cref{equ: x-ray imaging} to represent the illumination pattern without the object. This reference image may have artifacts, such as intensity dropoff towards the image boundaries and other non-uniformities in illumination. Additionally, scanner extrinsics and intrinsics may vary during scanning due to heat expansion and mechanical vibrations. Addressing these practical challenges will be the focus of future work.

\paragraph{Anisotropic effects}
Following existing CT reconstruction methods, we work under the isotropic assumption of X-ray imaging. However, in the real world, some X-ray transport effects, such as Compton scattering, are anisotropic. We do not explicitly model them but treat them as noises on X-ray projections. This is a necessary simplification for CT reconstruction but may induce inaccuracy for novel-view X-ray synthesis. Readers may refer to~\cite{cai2024radiative,gao2024ddgs} for using 3DGS for X-ray view synthesis.

\section{Broader impacts}
\label{sec: broader impact}
\paragraph{Impacts on real-world applications} Computed tomography is an essential imaging technique that is widely used in fields including medicine, biology, industry, etc. Our R$^2$-Gaussian enjoys superior reconstruction performance and fast convergence speed, making it promising to be implemented in real-world applications such as medical diagnosis and industrial inspection.

\paragraph{Impacts on research community} We discover a previously unknown integration bias problem in currently popular 3DGS. we speculate that this problem could be universal across all 3DGS-related works. Therefore, our rectification technique may apply to wider practical domains, not limited to tomography but also other tasks such as magnetic resonance imaging (MRI) reconstruction and volumetric-based surface reconstruction.

\end{document}